\documentstyle[prl,aps]{revtex}
\twocolumn
\begin{document}
\title{Comment on ``Searching for Evolutions of Pure States
into Mixed States in the Two-State System $K^0$-$\overline{K^0}$''}
\author{}
\author{
Fabio Benatti
}
\address{
Dipartimento di Fisica Teorica, Universit\`a di Trieste\\
Strada Costiera 11, 34014 Trieste, Italy\\
and\\
Istituto Nazionale di Fisica Nucleare, Sezione di Trieste
}
\author{
Roberto Floreanini}
\address{
Istituto Nazionale di Fisica Nucleare, Sezione di Trieste\\
Dipartimento di Fisica Teorica, Universit\`a di Trieste\\
Strada Costiera 11, 34014 Trieste, Italy
}
\author{}
\author{\\[7pt]
\small\parbox{14cm}{\small 
It has been recently proposed to study generic dynamical
evolutions of the neutral kaon system that go beyond
quantum mechanics. We explicitly show that, unless the condition of complete
positivity is enforced, those dynamics are physically inconsistent
and should be rejected.
\\
[7pt]{PACS numbers: 03.65.Ca, 13.20.Eb}}\vspace{-12pt}}
\author{}
\author{}
\maketitle

\narrowtext

In a recent paper, H.-J. Gerber discusses possible extensions of the
time-evolution of the neutral kaon system, beyond standard quantum mechanics.
The system can be modeled by a two-dimensional Hilbert space and
its states can be described by a $2\times2$ density matrix $\rho$.
The time-evolution equation considered in [1] is:
\begin{equation}
\label{1}
\frac{\partial}{\partial t}\rho(t)=L[\rho(t)]\ ,
\end{equation}
where $L$ is a generic linear map.
In this note, we would like to point out that time-evolutions generated
by equations of the form (\ref{1}), without further restrictions,
are in general physically inconsistent.

In going beyond conventional quantum mechanics, 
one has to be careful not to destroy the probabilistic interpretation of 
$\rho$, on which all physical considerations are based. To be specific,
the finite time evolution map $\gamma_t:\ \rho(0)\mapsto\, \rho(t)$
must at least preserve the positivity of the eigenvalues
of $\rho(t)$ for all times $t\geq0$. 
However, this condition alone does not guarantee 
the positivity of the eigenvalues of density matrices 
of correlated kaons evolving with $\gamma_t\otimes\gamma_t$: 
in fact, $\gamma_t$ has to be not only positive, but also
completely positive.\cite{2,3,5}
This requirement is physically unavoidable, since correlated kaons
are indeed produced in the decay of $\phi$ mesons, and are the focus
of experimental investigations in the so called $\phi$-factories.
Without the complete positivity of $\gamma_t$, 
the time evolution $\gamma_t\otimes\gamma_t$ of correlated kaons 
would be beset by negative eigenvalues of evolving
density matrices.\cite{2,3}

In general, since the neutral kaon system is unstable, the evolution
$\gamma_t$ must satisfy a forward in time composition law, 
decrease the trace of the kaon state and increase its von Neumann entropy.
Together with complete positivity, this fixes 
the generator $L$ in (\ref{1}) to be\cite{2,6}
\begin{equation}
\label{2}
L[\rho]=-i\left[M,\rho(t)\right]
        -{1\over2}\left({\mit\Gamma}\rho+\rho{\mit\Gamma}\right)
		+L_D[\rho]\ .
\end{equation}		
The first two pieces in the r.h.s. of (\ref{2})
correspond to the Weisskopf-Wigner standard evolution with effective
hamiltonian $H=M-(i/2){\mit\Gamma}$. The term $L_D$ is a dissipative
contribution\cite{6}.
By writing $\rho$ as a $4$-vector with components
$(\rho_0,\rho_1,\rho_2,\rho_3)$ along the identity 
matrix $\sigma_0$ and the Pauli matrices $\sigma_i$, $i=1,2,3$,
$[L_D]$ acts as:
\begin{equation}
\label{3}
[L_D]=-2\left(\begin{array}{cccc}
0&0&0&0\\
0&a&b&c\\
0&b&\alpha&\beta\\
0&c&\beta&\gamma
\end{array}
\right)\ ,
\end{equation}
and the six real parameters must satisfy
\begin{eqnarray}
\nonumber
&&2R\equiv\alpha+\gamma-a\geq0\ ,\qquad RS\geq b^2\ ,\\
\nonumber
&&2S\equiv a+\gamma-\alpha\geq0\ ,\qquad RT\geq c^2\ ,\\
\nonumber
&&2T\equiv a+\alpha-\gamma\geq0\ , \qquad ST\geq\beta^2\ ,\\
\nonumber
&&RST\geq 2\, bc\beta+R\beta^2+S c^2+T b^2\ .\\
\nonumber
\end{eqnarray}
\vskip -.5cm

Concretely, let $\rho_\pm=(\sigma_0\pm\sigma_3)/2$, 
$\sigma_\pm=(\sigma_1\pm i\sigma_2)/2$ and set for simplicity
$M={\mit\Gamma}=0$, 
$a=b=c=0$, $\alpha=\gamma=1$, $\beta=1/2$:
since $0=ST<\beta^2=1/4$, $\gamma_t$ is positive but not
completely positive. The initial projections $\rho_\pm$ evolve into
$\rho_\pm(t)=1/2(\sigma_0\mp s\sigma_2 \pm r\sigma_3)$, 
where $r=e^{-2t}\cosh{t}$ and $s=e^{-2t}\sinh{t}$, and keep
positive eigenvalues, while $\sigma_\pm(t)=1/2(\sigma_1\pm ir\sigma_2
\mp is\sigma_3)$. Therefore, the initial state
$\rho_S=1/2(\rho_+ \otimes \rho_- + \rho_- \otimes \rho_+ - 
  \sigma_+ \otimes \sigma_- - \sigma_-\otimes \sigma_+)\ ,$
describing two correlated kaons,
evolves with $\gamma_t\otimes\gamma_t$ into
$$
  \rho_S(t)={1\over 4}\pmatrix{1-x&-iy&-iy&-1+x\cr 
                        iy&1+x&-1-x&iy\cr
                        iy&-1-x&1+x&iy\cr 
                        -1+x&-iy&-iy&1-x}\ ,
$$
where $x=e^{-4t}\cosh(2t)$, $y=e^{-4t}\sinh(2t)$, and develops a negative 
eigenvalue as soon as $t>0$.\cite{3,5}

Concluding, physical consistency demands that it is the time evolution 
generated by the equation in (\ref{2}), 
and not the generic one given by (\ref{1})
that should be compared with the experimental data.\cite{4,5}

\end{document}